# Orbital angular momentum photonic quantum interface

Zhi-Yuan Zhou,[1,2,#] Yan Li,[1,2,#] Dong-Sheng Ding,[1,2] Wei Zhang,[1,2] Shuai Shi,[1,2] Bao-Sen Shi,[1,2*] and Guang-Can Guo[1,2]

[1]*Key Laboratory of Quantum Information, University of Science and Technology of China, Hefei, Anhui 230026, China*

[2]*Synergetic Innovation Center of Quantum Information & Quantum Physics, University of Science and Technology of China, Hefei, Anhui 230026, China*

[*]*Corresponding author: drshi@ustc.edu.cn*

Light carrying orbital angular momentum (OAM) has great potential in enhancing the information channel capacity in both classical and quantum optical communications. Long distance optical communication requires the wavelengths of light are situated in the low-loss communication windows, but most quantum memories currently being developed for use in a quantum repeater work at different wavelengths, so a quantum interface to bridge the wavelength gap is necessary. So far, such an interface for OAM-carried light has not been realized yet. Here, we report the first experimental realization of a quantum interface for a heralded single photon carrying OAM using a nonlinear crystal in an optical cavity. The spatial structures of input and output photons exhibit strong similarity. More importantly, single-photon coherence is preserved during up-conversion as demonstrated.

**Keywords:** sum frequency generation; orbital angular momentum; frequency conversion; spontaneous parametric down conversion

## INTRODUCTION

Photons are very important information carriers for transferring quantum states between remote physical systems, such as atomic ensembles, ions and solid-state systems[1–7] acting as quantum memories[5-9] and quantum information processors.[10] Light carrying orbital angular momentum (OAM) has stimulated considerable research interest in both classical and quantum optical fields, has exciting applications, including optical manipulation and trapping,[11,12] high-precision optical measurements,[13–15] high-capacity free-space and fiber optic communications,[16, 17] and studies of fundamental quantum physics.[18-21] In quantum communication, due to the inherent infinite dimension of OAM, photons encoded in OAM space can significantly increase the information channel capability in quantum key distribution.[22–24] A photon in telecom band or in free-space communication window is vital to construct a long-distance high-capacity quantum communication network. So far, most quantum memories operate in the visible wavelength range, [5-9] only few memories can work in telecom band.[25] Furthermore, the signal stored is an attenuated coherent light and has the Gaussian mode. Only recently, the storage of telecom wavelength entanglement is realized in an erbium-doped optical fibre,[26] but the spatial mode used is Gaussian mode. Quantum memories for photons with OAM have recently been realized, [27, 28] but all work in visible range. So a quantum interface to bridge the wavelength gap is necessary. So far, such an interface for OAM-carried light has not been realized yet.

There are some experimental realizations of quantum interfaces for single photons with Gaussian shapes,[29–38] either by using second-order nonlinear processes in nonlinear crystals or by third-order nonlinear processes in atomic ensembles.[39, 40] Frequency conversion using nonlinear crystals is much more attractive for practical applications because it can offer wide phase matching wavelength range, in contrast to using atomic ensembles. Most of previous experiments used periodically-poled LiNbO$_3$ (PPLN) bulk crystals or waveguides to perform the frequency conversion, near unity conversion efficiency can be reached in waveguide PPLN crystals and the quantum properties of the single photons are preserved. The frequency conversion of photons with OAM using waveguide crystals is not possible because OAM modes cannot propagate in waveguides. However, our recent studies on the frequency conversion of OAM-carried light offer the possibility for realizing this aim with bulk periodically-poled nonlinear crystals.[41–43]

In this work, we report the first experimental realization of an OAM photonic quantum interface by up-converting a heralded OAM-carried single photons from 1560 nm to 525 nm using the cavity-enhanced sum frequency generation (SFG). The conversion efficiency can reach 8% for photons carrying OAM of $1\hbar$. We clearly demonstrate that the spatial structure of input and output photons exhibit strong similarity. We also show that the coherence properties of the single photons are retained in the conversion process. This primary study will pave the way for high-dimensional quantum information processing, creating a link between different quantum systems that work in different wavelengths by using OAM degree of freedoms of photons.

[#] **These two authors have contributed equally to this article.**

## MATERIALS AND METHODS

### Details of the SPDC and SFG crystals

Both the spontaneous parametric down conversion (SPDC) crystal and the SFG crystal are periodically poled potassium titanyl phosphate (PPKTP), which are manufactured by Raicol Crystals, and all of these crystals have dimensions of 1 mm×2 mm×10 mm. The type-II SPDC crystal has a poling period of 46.2 μm; both end faces of the crystal are anti-reflection coated for 780 nm and 1560 nm, and the measured quasi-phase matching temperature of the crystal is 23.6°C. The type-I SFG crystal has a poling period of 9.375 μm; both end faces of the crystal are anti-reflection coated for 525 nm, 795 nm and 1560 nm, and the measured quasi-phasing matching temperature of the crystal is 39.4°C. The 795 nm wavelength corresponding to Rb$^{85}$ D1 line, the 1560 nm is at telecom band suitable for long distance transmission. The SFG beam can be used to generate two-color signal and idler photon source at 795 nm and 1560 nm in another crystal which has the same parameter as the SFG crystal.

### SFG cavity design

The bow-tie ring cavity is designed for a single resonance at 795 nm, and the total cavity length is 547 mm. The input coupling mirror M1 has transmittance of 3% at 795 nm. Mirror M2 is highly reflectively coated at 795 nm (R >99.9%), and a piezoelectric element (PZT) is attached to it to scan and lock the cavity. The two concave mirrors, M3 and M4, both have curvatures of 80 mm; M3 has a high transmittance coating for 1560 nm (T>99%) and is highly reflectively coated for 795 nm (R>99.9%), while M4 has a high transmittance coating for 525 nm (T>98%) and is highly reflectively coated for 1560 nm and 795 nm (R>99.9%). The fundamental cavity mode has a beam waist of 33 μm at the mid-points of mirrors M3 and M4.

## RESULTS AND DISCUSSION

***Theoretical model.*** The quantum theory for SFG of continuous waves in second-order nonlinear crystals is shown as follows. Three waves are involved in the up-conversion process: one strong pump beam at frequency $\omega_p$, one signal beam to be converted at frequency $\omega_s$, and the up-converted beam at frequency $\omega_{SFG}$, where the frequencies of the interacting waves satisfy $\omega_{SFG} = \omega_p + \omega_s$. The entire conversion process can be described by the following Hamiltonian:[44]

$$H_I = i\hbar\kappa(\hat{a}_s \hat{a}_{SFG}^\dagger - \hat{a}_s^\dagger \hat{a}_{SFG}), \quad (1)$$

Here, $\hat{a}_s$ and $\hat{a}_{SFG}$ are the annihilation operators for the signal and the up-converted photons, respectively. $\kappa = gE_p$ is a constant, where $g$ is proportional to the second-order susceptibility $\chi^{(2)}$; $E_p$ denotes the pump beam's electrical field amplitude. The Heisenberg equations of motion in the interaction picture are

$$\frac{d\hat{a}_s}{dt} = -\kappa \hat{a}_{SFG}, \quad (2)$$

$$\frac{d\hat{a}_{SFG}}{dt} = \kappa \hat{a}_s, \quad (3)$$

The solutions to these two equations for a nonlinear interaction length $L$ are given by:

$$\hat{a}_s(L) = \cos(\kappa L)\hat{a}_s(0) - \sin(\kappa L)\hat{a}_{SFG}(0), \quad (4)$$

$$\hat{a}_{SFG}(L) = \sin(\kappa L)\hat{a}_s(0) + \cos(\kappa L)\hat{a}_{SFG}(0), \quad (5)$$

The single photon up-conversion efficiency is defined as $\eta = N_{SFG}(L)/N_s(0)$, where $N_i = \langle \hat{a}_i^\dagger \hat{a}_i \rangle_T$ is the mean photon number in the measurement time $T$; $\hat{a}_s(0)$ and $\hat{a}_{SFG}(0)$ are the annihilation operators for signal and SFG photons at 0 interaction length (at the input face of the crystal), respectively. $N_s(0)$ is the input signal photon number at front face of the crystal. For a practical up-converter, $N_{SFG}(0) = 0$, and $\eta = \sin^2(\kappa L)$. The perfect conversion is achieved under the condition of $\kappa L = \pi/2$. While the theoretical model above is for a Gaussian mode photon, it can naturally be generalized for a photon with OAM.

***Conversion efficiencies for different OAMs.*** For frequency up-conversion using two Gaussian light beams, the quantum conversion efficiency of the signal light can be expressed as [45]

$$\eta = \sin^2\left(\frac{\pi}{2}\sqrt{\frac{P}{P_{max}}}\right), \quad (6)$$

Here, $P$ is the circulating power of the pump beam in cavity, and $P_{max}$ is the pump power that gives unity conversion efficiency. The expression for $P_{max}$ is:

$$P_{max} = \frac{\varepsilon_0 c n_s n_{SFG} \lambda_p \lambda_s \lambda_{SFG}}{16\pi^2 d_{eff}^2 L h(0,\xi)}, \quad (7)$$

Here, $\varepsilon_0$ and $c$ represent the vacuum permittivity and the speed of light in a vacuum; $n_s$ and $n_{SFG}$ are the refractive indices of the signal and up-converted beams; $\lambda_s$, $\lambda_{SFG}$ and

$\lambda_p$ are the wavelengths of the three interaction waves; $d_{eff}$ is the effective nonlinear coefficient; $L$ is the crystal length; and $h(\xi)$ is a parameter dependent on the focusing parameter $\xi$, please refer to the supplementary of information for details.

In the situation where the pump beam is in a Gaussian mode and the signal beam carries OAM, the SFG power is calculated as

$$P_{SFG} = \frac{16\pi^2 d_{eff}^2 P P_s L 2^l}{\varepsilon_0 c n_S n_{SFG} \lambda_{SFG}^2 \lambda_p} h(l,\xi) \quad , \qquad (8)$$

where $P_s$ is the signal power, and $l$ is the OAM index of the signal. For $l=0$, equation (8) is reduced to the SFG with two Gaussian beams. For more detailed derivations of equations (7) and (8), please refer to the Supplementary of information.

***Up-conversion of a classical light with OAM.*** To obtain an overview of frequency up-conversion of OAM-carried light, we first perform an experiment using coherent light. We want to mention the fact that demonstrations of OAM frequency conversion and conservation using classical light are also widely studied in birefringence phase matching crystals [46-49]. The experimental setup is shown in Figure 1(a). High conversion efficiency can be achieved by placing a periodically-poled KTiOPO$_4$ (PPKTP) crystal inside a ring cavity (please refer to materials and methods section for details of the PPKTP crystal and the cavity design). The strong pump beam is provided by a Ti:sapphire laser (Coherent, MBR110), and the light carrying OAM by a vortex phase plate (VPP, RP Photonics) for conversion comes from a diode laser (Toptica, pro design). The cavity is actively locked using the Hansch-Couillaud technique[50]. We measure the SFG power versus the signal power for various OAM with a fixed pump power of 750 mW, the results are shown in Figure 2(a). We conclude that the SFG power is linearly proportional to the input signal power for OAM values $l$ of 0, 1 and 2; the images inserted across the lines are the corresponding spatial shapes for the various OAM modes, and these images are acquired using a charge-coupled device (CCD) camera. The power conversion efficiencies determined using $\eta_{power} = P_{525}/P_{1560}$ are 0.66, 0.259 and 0.0893 for $l$ values of 0, 1 and 2, respectively. The corresponding quantum conversion efficiencies defined by $\eta_{quantum} = \eta_{power} \lambda_{525}/\lambda_{1560}$ are 0.224, 0.0833 and 0.0296, respectively. The conversion efficiencies will keep unchanged against the signal power according to the linearity of the SFG process. We also calculate the quantum conversion efficiency for different OAMs, the results are showed in Figure 2(b), where the efficiencies are normalized with respect to the Gaussian mode. The theoretical predictions are well in agreement with experimental results. The differences in the conversion efficiencies for different OAMs are mainly caused by different overlaps between the signal and the pump beams, this can be explained by equation (8) (for further details, please refer to the Supplementary of information). Differences in conversion efficiency for different OAM modes could be somehow compromised by pre-engineering the focus parameter and amplitude of the input signal beam.

We then test our system with attenuated coherent light. The results are shown in Figure 2(c)–(f), which are obtained using a single-photon-counting camera (Andor, ICCD) by setting it in fire only mode, each image is accumulated with 360 frames and the background is subtracted (the dark count is 600 for each pixel in each frame), the exposure time of the ICCD is set to be 1 s. The numbers of frames required for summation to obtain the final images shown in Figure 2(c)–(f) are (6, 7, 5, 7) *360, respectively. The numbers of input photons are calibrated using an InGaAs single photon avalanche detector (APD) (Lightwave, Princeton, 30 MHz trigger rate, 1 ns detection window). The recorded count rates by APD are 11.7 k/s, 21.2 k/s and 16.8 k/s and 21.2 k/s for photons in OAM state of $|1\rangle$, $|2\rangle$, $1/\sqrt{2}(|1\rangle+|-1\rangle)$ and $1/\sqrt{2}(|2\rangle+|-2\rangle)$, respectively. The actual photon number rates at the crystal's input face are 2.1M/s, 3.7M/s, 2.8M/s and 3.1 M/s after we consider the losses of the VPP and the input mirror M3 (the transmission efficiencies for the four input states measured with strong coherent laser beams are 0.80, 0.79, 0.74 and 0.67 respectively), the duty cycle of the APD (1/33.33) and detection efficiency of 0.15 per gate. The typical donut structures can be clearly distinguished for the different OAM-carried input beams in Figure 2(c) and (d), the theoretical predictions are showed in figure 2(g)-(h).

In addition to up-conversion of light with single OAM, we also perform up-conversion of light with OAM superposition. A modified Sagnac interferometer[20, 27, 41] is used to generate the OAM superposition, the generated superposition state is

$$|\Phi\rangle_{in} = \frac{1}{\sqrt{2}}\left(|H\rangle|l\rangle + e^{i\theta}|V\rangle|-l\rangle\right), \qquad (9)$$

where $|H\rangle$ and $|V\rangle$ denote the polarization of the signal beam, $|l\rangle$ represents the OAM of the beam and $\theta$ is a phase dependent on the position of the half wave plate (HWP) at the input port of the interferometer. When we insert another HWP with optical axes placed at 22.5° relative to the horizontal direction, the up-converted SFG light state is (for details of the derivation, see supplementary of information )

$$|\Phi\rangle_{out} = \frac{1}{\sqrt{2}}(|l\rangle + e^{i\theta}|-l\rangle)|V\rangle, \qquad (10)$$

Equation (10) is obtained by projection 45 degree rotated two polarization components in equation (9) onto the vertical direction, it shows that the up-converted state is a superposition of two OAMs with the same absolute value but opposite sign, the interference pattern between them has $2l$ maximum in the azimutal direction, similar to $2l$ "petal". The experimental results for $l=1, 2$ are shown in Figure 2(e) and (f), typical petals in the interference patterns show good agreement with the theoretical expectations (Figure 2(i) and (j)). The mean photon number of the attenuated light at the input face of the SFG crystal in 1ns detection window is about 0.002, which is in the same order with a typical SPDC photon source, therefore, our system is capable of converting OAM-carried photons from SPDC.

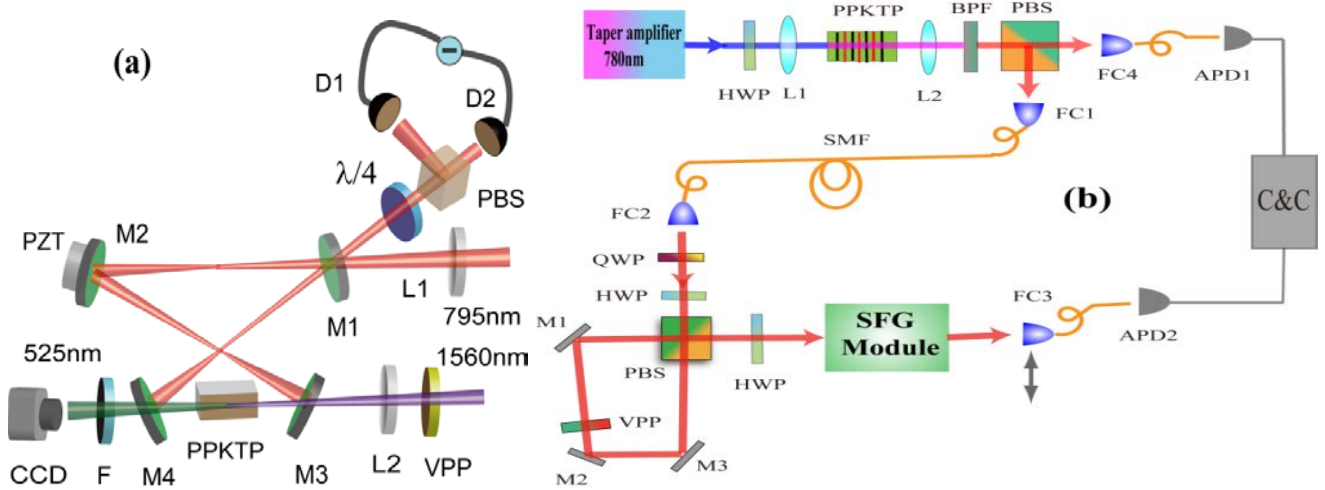

**Figure 1 Setup for the cavity-enhanced up-converter module (a) and for up-converting a herald single photon with OAM (b).** L1, L2: lenses; M1-M4: cavity mirrors; VPP: vortex phase plate; PBS: polarizing beam splitter; F: filters; PPKTP: periodically-poled KTP crystal. HWP, QWP: half (quarter) wave plate; BPF: band pass filter; FC1-FC4: fiber couplers; SMF: single mode fiber; APD1(APD2): InGaAs (silicon) avalanche detector.

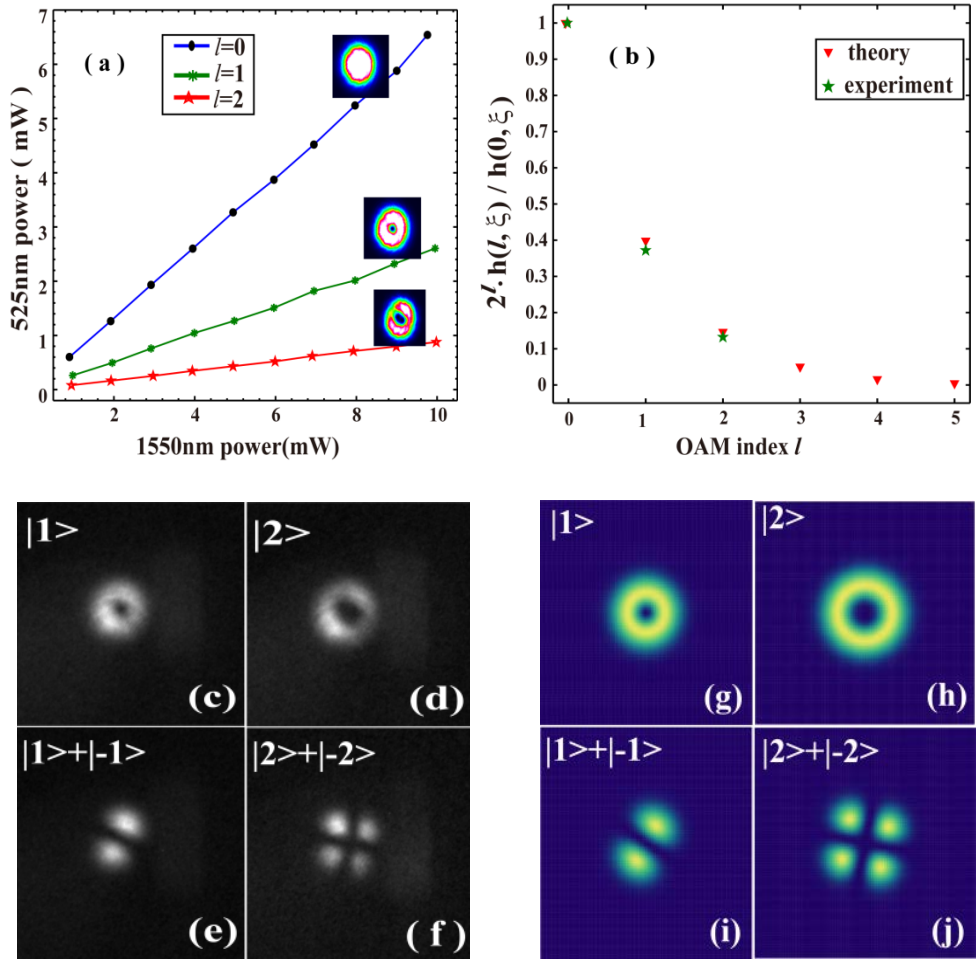

**Figure 2 Experimental results with strong and attenuated coherent light at single photon level respectively.** (a) The lines show the relationships between the input signal power and the SFG output powers for *l*=0, 1, and 2, respectively. The inserted images across the lines are the spatial shapes for the corresponding SFG light; (b) Experimental results and theoretical simulations of up-conversion efficiency for different OAM based on equation (8); (c)-(f) show the up-converted images of light with single OAM and superpositions input of *l*=1, 2, respectively; (g)-(i) are the corresponding theoretical simulation results for (c)-(f) respectively.

***Conversion of a heralded single photon with OAM.*** The photon pair is prepared by using a 780-nm laser with 120 mW power to pump a type-II PPKTP crystal, generating degenerate signal and idler photons at 1560 nm. The non-classical nature of the signal and idler photons can be characterized using the intensity cross-correlation $g_{s,i}^{(2)}$ between them.[5, 51] The normalized second-order correlation function are defined as

$$g_{jk}^{(2)}(\tau) = \frac{\langle E_j^\dagger(t) E_k^\dagger(t+\tau) E_k(t+\tau) E_j(t) \rangle}{\langle E_j^\dagger(t) E_j(t) \rangle \langle E_k^\dagger(t+\tau) E_k(t+\tau) \rangle}, \quad (11)$$

Where indices $j, k \in \{s, i\}$ represent the signal or idler photon, respectively. The measurements of $g_{j,k}^{(2)}(\tau)$ consists of first determining the rate of coincidence detections between mode $j$ and $k$ at a time delay $\tau$. This is effectively a measurement of the non-normalized second-order coherence function, which is the numerator in equation (11) The normalization is then performed with respect to the rate of coincidences between photons from uncorrelated pairs created at times differing by much more than the coherence time of the photons.

If we assume that the auto-correlations $1 \leq g_{s,s}^{(2)}, g_{i,i}^{(2)} \leq 2$ are satisfied, then non-classicality is provided for measured cross-correlation $g_{s,i}^{(2)} > 2$. Experimental we obtain $g_{s,i}^{(2)} = 162$ between the input signal and idler photons. The crystal is described in detail in the supplementary materials and the performance of the crystal is described in our previous works.[52–54] The experimental setup for up-converting a herald single photon with OAM is shown in Figure 1(b). We first measure the spatial structure of the up-converted photon by ICCD. The results are shown in Figure 3, where Figure 3(a) and (b) show the results for the OAM states $|1\rangle$ and $|2\rangle$, and Figure 3(c) and (d) show the results for the superposition states $1/\sqrt{2}(|1\rangle+|-1\rangle)$ and $1/\sqrt{2}(|2\rangle+|-2\rangle)$, respectively. Images in Figure 3(a)-(d) show that the photon with both single OAM value or the OAM superposition can be up-converted, as the typical donut-shapes and interference patterns are clearly distinguished. In this experiment, the ICCD has the same settings as used in the previous experiments with the attenuated coherent light.

We also measure the cross-correlations for different input OAM states by coupling the up-converted photons into a single mode fiber (SMF). We first perform coincidence measurements between the idler and the up-converted signal photons with the input signal photon in the Gaussian mode. The results are shown in Figure 3(e); The measured cross-correlation in 2 ns coincidence window is $g_{s,i}^{(2)} = 25$, which demonstrates that the up-converted signal photon and the idler photon are in non-classical correlation. We then perform coincidence measurements for signal photons with OAM by scanning the SMF in the horizontal direction. The results are shown in Figure 3(f)–(i). Figure 3(f) shows the result for a strong coherent input beam with $l=1$, where the position-dependent power coupling into the SMF has a dip at the center, and the theoretical fit shows good agreement with the experimental data; Figure 3(g) shows the measured $g_{s,i}^{(2)}$ for a single photon input, where the $g_{s,i}^{(2)}$ has a dip in the raw data, and the maximum $g_{s,i}^{(2)}$ is 15.5. The deviation of the experimental data from the theoretical fit at the center is a result of the impurity of the up-converted OAM mode, which is mainly caused by misalignment and spontaneous Raman scattering noise, the noises in the up-conversion process is discussed in the supplementary of materials. Figure 3(h) and (i) show the results for $l=2$, and the maximum $g_{s,i}^{(2)}$ for single photon input is 6.5. $g_{s,i}^{(2)} > 2$ for $l=1,2$ are clear evidence proving the existed non-classical correlation in spatial shape between the up-converted signal and the idler photons. We should point out that the filtering of the up-conversion spatial mode using SMF introduces loss in the detection, if the spatial mode is detected directly, the value of the cross-correlation will be even larger. To show that the coherence properties are retained in the conversion process, we let the input signal photon be in the state of Equation (9), then the up-converted signal photon state is in the form of Equation (10). By filtering out a single petal of the interference pattern use a pinhole and coupling it into the SMF, we measure the coincidence dependent on the phase $\theta$ through rotating the HWP at the input port of the interferometer. This method is introduced in refs.[20, 27] The filtering operation can be described using the following projection operator $\hat{P} = |\Pi\rangle\langle\Pi|$, where $|\Pi\rangle = 1/\sqrt{2}(|l\rangle + e^{i\varphi}|-l\rangle)$, phase $\varphi$ represents the position of the pinhole, the coincidence rate is related to the following expression:

$$C(\theta, \varphi) \propto \langle \Phi|_{out} \hat{P} |\Phi\rangle_{out} = 4\cos^2(\frac{\theta - \varphi}{2}). \quad (12)$$

Equation (12) shows that the coincidence rate is a sinusoidal function of phase $\theta$. The results for $l=1,2$ are shown in Figure 3(j) and (k), where the corresponding visibilities are 89%±6% and 89%±7%, respectively, and the error bars are estimated by assuming Poisson statistics for the measured data. Usually, the quality of a quantum transforming process is characterized using fidelity of process. For up-conversion of quantum states, the fidelity is defined as $F=\langle\Phi|_{in}\rho_{out}|\Phi\rangle_{in}$,[55] where $\rho_{out}=V|\Phi\rangle_{in}\langle\Phi|_{in}+(1-V)/2$ is the output density matrix and V is the visibility of interference, the fidelity is related to the visibility as $F=(1+V)/2$. Therefore the fidelities of up-conversion process for $l=1,2$ are 0.94±0.03 and 0.95±0.04 respectively.

The effective up-conversion of OAM modes high than $l=2$ are also possible by optimizing the experimental parameters. The feasible methods are: (i) narrowing the spectral bandwidth of the signal photon; (ii) increasing the length of the SFG crystal; (iii) changing the cavity geometric dimension to increase the beam waist inside the crystal in order to support high order spatial mode. The present setup is possible for up-conversion of some simple images at ultra-weak power (pW level), such as lower order OAM modes, OAM superposition mode and simple spatial shapes with spatial symmetry.

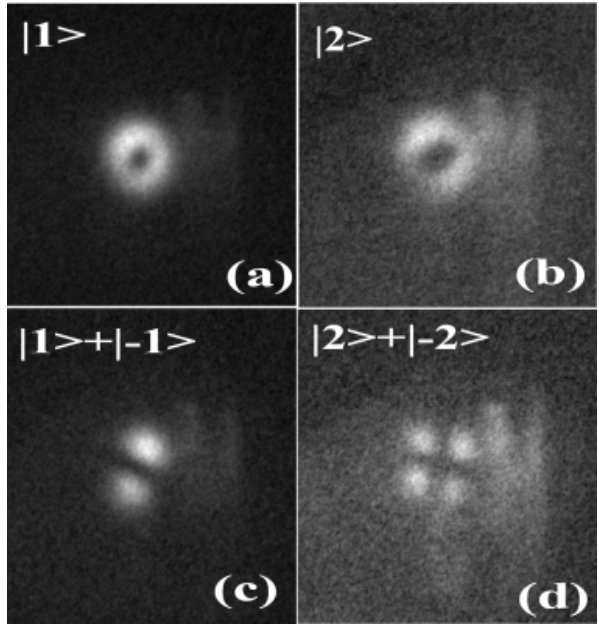

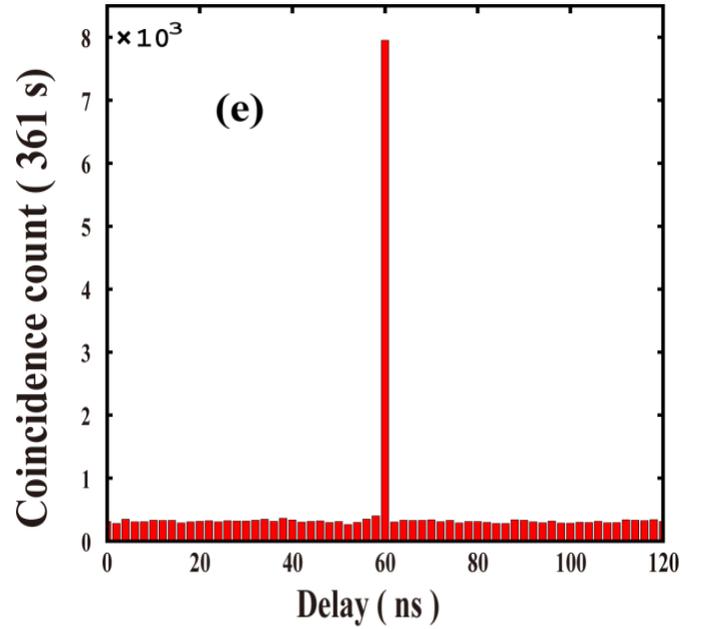

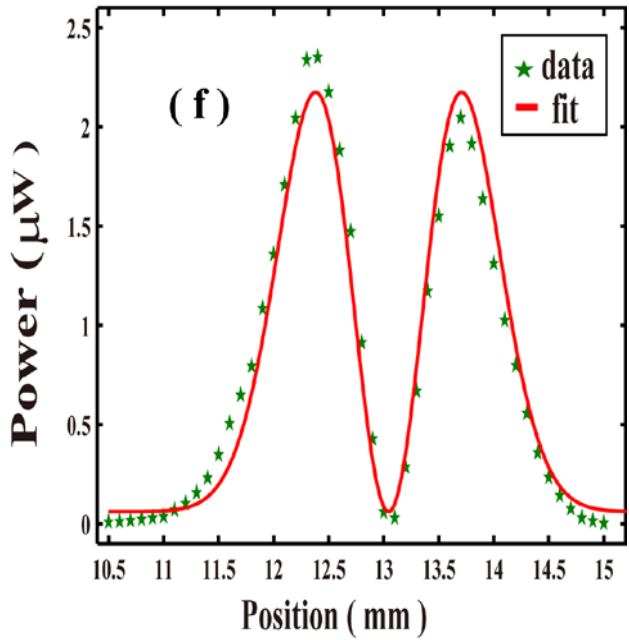

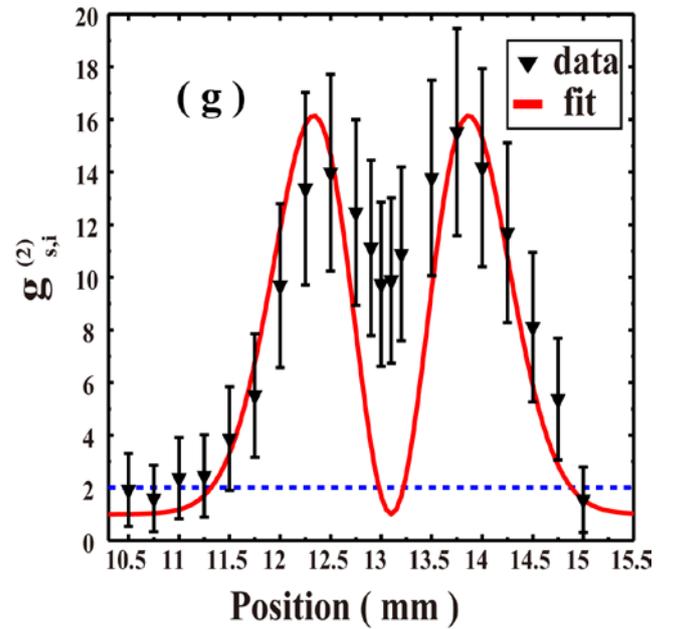

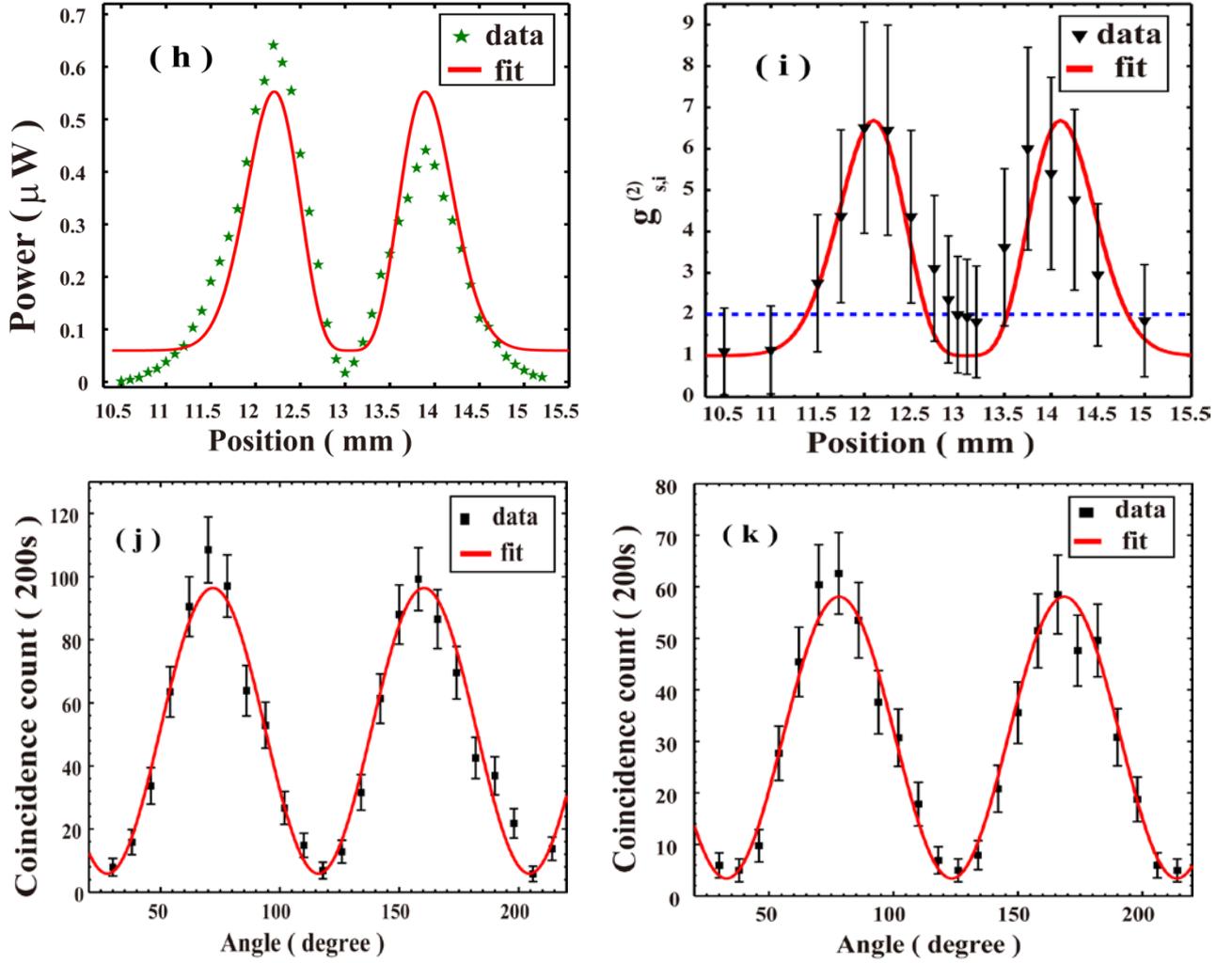

**Figure 3 Experimental results for heralded single photon from SPDC.** (a)–(d) show SFG photon images taken using the ICCD for different input states; (e) coincidence count between idler and up-converted signal photon with the Gaussian spatial shape; (f) and (h) show one-dimensional scanning results for position-dependent power for $l=1,2$, respectively, using strong coherent pump beams; (g) and (i) show corresponding $g_{s,i}^{(2)}$ measurement results for single photon signal inputs, error bars are estimated by assuming Poison statistics of photon measurements; (j) and (k) show the phase-dependent coincidence counts produced by rotating the HWP for $l=1,2$, respectively.

## CONCLUSIONS

We have realized an efficient photonic quantum interface for single photon with both single OAM and OAM superposition. The conversion efficiency for $l=1$ is 8.33% using our present setup and the coherence properties are retained in the up-conversion process. Also, the detailed theoretical description of OAM-carried light up-conversion provides a useful guide for optimizing the conversion process. This primary study will pave the way for high-dimensional quantum information processing in the OAM degree of photons, which create a link between different quantum systems that work in different wavelengths. The present setup can also be possibly used for single-photon-level image up-conversion detection with optimized experimental parameters, which will be of potential importance in many fields, such as biology, astrophysics, night-vision technology, chemical sensing.

## ACKNOWLEDGEMENTS

The authors would like to thank Dr. Bi-Heng Liu for loaning the InGaAs single photon detectors. This work was supported by the National Fundamental Research Program of China (2011CBA00200), the National Natural Science Foundation of China (11174271, 61275115, and 61435011) and the Innovation Fund from the Chinese Academy of Sciences.

## References

1. Blinov BB, Moehring DL, Duan LM, Monroe C. Observation of entanglement between a single trapped atom and a single photon. Nature 2004; **428**: 153-157.
2. Matsukevich DN, Kuzmich A. Quantum state transfer between matter and light. Science 2004; **306**: 663-666.
3. Togan E, Chu Y, Trifonov AS, Jiang L, Maze J, Childress L, *et al.* Quantum entanglement between an optical photon and a solid-state spin qubit. Nature 2010; **466:** 730-734.

4  Piro, N, Rohde F, Schuck C, Almendros M, Huwer J, Ghosh J, *et al.* Heralded single-photon absorption by a single atom. Nat. Phys. 2011; **7**: 17-20.

5  Clausen C, Usmani I, Bussieres F, Singouard N, Afzelius M, de Riedmatten H, Gisin N. Quantum storage of photonic entanglement in a crystal. Nature 2011; **469**: 508-511.

6  Bussieres F, Clausen C, Tiranov A, Korzh B, Verma VB, Nam SW, et. al. Quantum teleportation from a telecom-wavelength photon to a solid-state quantum memory. Nat. Photon. 2014; **8**:775-778.

7  Riedlander D, Kutluer K, Ledingham PM, Gundogan M, Fekete J, Mazzera M, et. al. Quantum storage of heralded single photons in a Praseodymium-doped crystal. Phys. Rev. Lett. 2014; **112**:040504.

8  Briegel HJ, D ü r W, Cirac JI, Zoller P. Quantum repeaters: the role of imperfect local operations in quantum communication. Phys. Rev. Lett. 1998; **81**: 5932-5935.

9  Duan LM, Lukin MD, Cirac JI, Zoller P. Long-distance quantum communication with atomic ensembles and linear optics. Nature 2001; **414**: 413-418.

10  Lloyd S, Shahriar MS, Shapiro JH, Hemmer PR. Long distance unconditional teleportation of atomic states via complete Bell state measurements. Phys. Rev. Lett. 2001; **87**: 167903.

11  Dholakia K, Cizmar T. Shaping the future of manipulation. Nat. photon. 2011; **5**:335-342.

12  Parterson L, Macdonald MP, Arlt J, Sibbett W, Bryant PE, Dholakia K. Controlled rotation of optically trapped microscopic particles. Science 2001; **292**:912-914.

13  D'Ambrosio V, Spagnolo N, Re LD, Slussarenko, S, Li Y, Kwek LC, *et al.* Photonics polarization gears for ultra-sensitive angular measurements. Nat. Commun. 2013; **4**: 2432.

14  Lavery MPJ, Speirits FC, Barnett SM, Padgett MJ. Detection of a spinning object using light's orbital angular momentum. Science 2013; **341**: 537-540.

15  Zhou ZY, Li Y, Ding DS, Zhang W, Shi S, Shi BS. Optical vortex beam based optical fan for high-precision optical measurements and optical switching. Opt. Lett. 2014; **39**: 5098-5101.

16  Wang J, Yang JY, Fazal IM, Ahmed N, Yan Y, Huang H, *et al.* Terabit free-space data transmission employing orbital angular momentum multiplexing. Nat. Photon. 2012; **6**: 488-496.

17 Bozinovic N, Yue Y, Ren YX, Tur M, Kristensen P, Huang H, *et al.* Terabit-scale orbital angular momentum mode division multiplexing in fibers. Science 2013; **340**:1545-1548.

18  Mair A, Vaziri A, Weihs G, Zeilinger A. Entanglement of the orbital angular momentum states of photons. Nature 2001; **412**:313-316.

19  Leach J, Jack B, Romero J, Jha AK, Yao AM, Franke-Arnold S, *et al.* Quantum correlations in optical angle-orbital angular momentum variables. Science 2010; **329**: 662-665.

20  Fickler R, Lapkiewicz R, Plick WN, Krenn M, Schaeff C, Ramelow S, *et al.* Quantum entanglement of high angular momenta. Science 2012; **338**: 640-643.

21  Dada AC, Leach J, Buller GS, Padgett MJ, Andersson E. Experimental high-dimensional two-photon entanglement and violations of generalized Bell inequalities. Nat. Phys. 2011; **7**: 677-680.

22  Barreiro JT, Wei TC, Kwiat PG. Beating the channel capacity limit for linear photonic superdense coding. Nat. Phys. 2008; **4**: 282 -286.

23  Bechmann-Pasquinucci H, Tittel W. Quantum cryptography using larger alphabets. Phys. Rev. A 2000, **61**: 062308.

24  Vallone G, D'Ambrosio V, Sponselli A, Slussarenko S, Marrucci L, Sciarrino F, *et al.* Free-space quantum key distribution by rotation-invariant twisted photons. Phys. Rev. Lett. 2014; **113**: 060503.

25  Lauritzen B, Minar J, de Riedmatten H, Afzelius M, Sangouard N, Simon C, *et. al.* Telecommunication-wavelength solid-state memory at the single photon level. Phys. Rev. Lett. 2010; **104**: 080502.

26  Saglamyurek E, Jin J, Verma VB, Shaw MD, Marsili F, Nam SW, *et al.* Quantum storage of entangled telecom-wavelength photons in an erbium-doped optical fibre. Nat. Photon. 2015; **9**: 83-87.

27  Ding DS, Zhou ZY, Shi BS, Guo GC. Single-photon-level quantum imaging memory based on cold atomic ensembles. Nat. Commun. 2013; **4**: 2527.

28  Nicolas A, Veissier L, Giner L, Giacobino E, Maxein D, Laurat J. A quantum memory for orbital angular momentum photonic qubits. Nat. Photon. 2014; **8**: 234-238.

29  Tanzilli S, Tittel W, Halder M, Alibart O, Baldi P, Gisin N, *et al.* A photonic quantum information interface. Nature 2005; **437**: 116 -120.

30  Takesue H. Single-photon frequency down-conversion experiment. Phys. Rev. A 2010; **82**: 013833.

31  Curtz N, Thew R, Simon C, Gisin N, Zbinden H. Coherent frequency down-conversion interface for quantum repeaters. Opt. Express 2010; **18**: 22099

-22104.

32  Zaske S, Lenhard A, Kebler CA, Kettler J, Hepp C, Arend C, *et al.* Visible-to-telecom quantum frequency conversion of light from a single quantum emitter. Phys. Rev. Lett. 2012; **109:** 147404.

33  Takesue H. Phys. Rev. Lett. Erasing distinguishability using quantum frequency up-conversion. 2008; **101:** 173901.

34  Rakher MT, Ma L, Slattery O, Tang X, Srinivasan K. Quantum transduction of telecommunications-band single photons from a quantum dot by frequency upconversion. Nat. photon. 2010; **4:** 786-791.

35  McGuinness HJ, Raymer MG, McKinstrie CJ, Radic S. Quantum frequency translation of single-photon states in a photonic crystal fiber. Phys. Rev. Lett. 2010; **105:** 093604.

36  Ates S, Agha I, Gulinatti, A, Reach I, Rakher MT, Badolato, *et al.* Two-photon interference using background-free quantum frequency conversion of single photons emitted by an InAs quantum dot. Phys. Rev. Lett. 2012; **109:**147405.

37  Ikuta R, Kusaka Y, Kitano T, Kato H, Yamamoto T, Koashi M, *et al.* Wide-band quantum interface for visible-to-telecommunication wavelength conversion. Nat. Commun. 2011; **2:** 537.

38  Guerrerio T, Martin A, Sanguinetti B, pelc JS, Langrock C, Fejer MM, *et al.* Nonlinear interaction between single photons. Phys. Rev. Lett. 2014; **113:** 173601.

39  Radnaev AG, Dudin YO, Zhao R, Jen HH, Jenkins SD, Kuzmich A, *et al.* A quantum memory with telecom-wavelength conversion. Nat. Phys. 2010; **6:** 894-899.

40  Dudin YO, Radnaev AG, Zhao R, Blumoff JZ, Kennedy TAB, Kuzmich A. Entanglement of light-shift compensated atomic spin waves with telecom light. Phys. Rev. Lett. 2010; **105:** 260502.

41  Zhou ZY, Ding DS, Jiang YK, Li Y, Shi S, Wang XS, *et al.* Orbital angular momentum light frequency conversion and interference with quasi-phase matching crystals. Opt. Express 2014; **22:** 20298-20310.

42  Zhou ZY, Li Y, Ding DS, Zhang W, Shi S, Shi BS, *et al.* Highly efficient second harmonic generation of a light carrying orbital angular momentum in an external cavity. Opt. Express 2014; **22:** 23673-23678.

43  Li Y, Zhou ZY, Ding DS, Shi BS. Sum frequency generation with two orbital angular momentum carrying laser beams. J. Opt. Soc. Am. B 2015; **32**: 407-411.

44  Kumar P. Quantum frequency conversion. Opt. Lett. 1990; **15:**1476-1478.

45  Albota MA, Wong FNC. Efficient single-photon counting at 1.55 μm by means of frequency upconversion. Opt. Lett. 2004; **29:** 1449-1451.

46  Berzanskis A, Matijosius A, Piskarskas A, Smilgevicius V, Stabinis A. Conversion of topological charge of optical vortices in a parametric frequency converter. Opt. Commun. 1997; **140:** 273–276.

47  Cætano DP, Almeida MP, Ribeiro PHS, Hugenin JAO, dos Santos BC, Khoury AZ. Conservation of orbital angular momentum in stimulated down-conversion. Phys. Rev. A 2002; **66:** 041801.

48  Molina-Terriza G, Torner L, Minardi S, Trapani PD. Simultaneous frequency conversion and beam shaping for optical-tweezers applications. J. Mod. Opt. 2003; **50:** 1563-1572.

49  Devaux F, Passier R. Phase sensitive parametric amplification of optical vortex beams. Eur. Phys. J. D 2007; **42:** 133-137.

50  Hansch T, Couillaud B. Laser frequency stabilization by polarization spectroscopy of a reflecting reference cavity. Opt. Commun. 1980; **35:** 441-444.

51  Kuzmich A, Bowen WP, Boozer AD, Boca A, Chou CW, Duan LM, Kimble HJ. Generation of nonclassical photon pairs for scalable quantum communication with atomic ensembles. Nature 2003; **423:** 731-734.

52  Zhou ZY, Jiang YK, Ding DS, Shi BS. An ultra-broadband continuously-tunable polarization entangled photon-pair source covering the C+L telecom bands based on a single type-II PPKTP crystal. J. Mod. Opt. 2013; **60:** 720-725.

53  Zhou ZY, Jiang YK, Ding DS, Shi BS, Guo GC. Actively switchable nondegenerate polarization entangled photon pair distribution in dense wave-division multiplexing. Phys. Rev. A 2013; **87:** 045806.

54  Zhou ZY, Ding DS, Li Y, Wang FY, Shi BS. Cavity-enhanced bright photon pairs at telecom wavelengths with a triple-resonance configuration. J. Opt. Soc. Am. B 2014; **31:**128-134.

55  Marcikic I, de Riedmatten H, Tittel W, Zbinden H, Gisin N. Long-distance teleportation of qubits at telecommunication wavelengths. Nature 2003; **421:**509-5.

# Supplementary of materials

**Calculation of the conversion efficiency for different OAM modes**

The electrical fields of the three interacting light beams are defined as

$$\tilde{E}(r,z,t) = E(r,z,t)\exp[i(kz-\omega t)] + E^*(r,z,t)\exp[-i(kz-\omega t)]. \quad (S1)$$

For pump power $P$ in the Gaussian mode and signal power $P_s$ in the OAM mode with OAM index $l$, the expressions for the electrical amplitudes are:

$$E_p(r,z) = \sqrt{\frac{P}{\pi\varepsilon_0 n_p c}} \frac{1}{\omega_{0p}(1+\frac{iz}{Z_{0p}})} \exp\left[-\frac{r^2}{\omega_{0p}^2(1+\frac{iz}{Z_{0p}})}\right], \quad (S2)$$

$$E_s(r,z) = \sqrt{\frac{P_s}{\pi l!\varepsilon_0 n_s c}} \frac{(\sqrt{2}r)^l}{[\omega_{0s}(1+\frac{iz}{Z_{0s}})]^{l+1}} \exp\left[-\frac{r^2}{\omega_{0s}^2(1+\frac{iz}{Z_{0s}})}\right]\exp(il\varphi), \quad (S3)$$

where $n_i (i=p,s)$ are the refractive indexes of the pump and the signal beams inside the crystal, $\omega_{0i}(i=p,s)$ are the beam waists, and $Z_{0i} = \pi n_i \omega_{0i}^2/\lambda_i$ $(i=p,s)$ are the Rayleigh ranges of these beams, the term $\exp(il\varphi)$ is responsible for OAM conservation in the conversion process, and $\varphi = \tan^{-1}(\frac{y}{x})$.

The SFG in the PPKTP crystal can be described by the following coupling wave equations:

$$\begin{cases} \frac{dE_p}{dz} = \frac{2id_{eff}\omega_p}{n_p c} E_{SFG} E_s^* \exp(-i\Delta kz) \\ \frac{dE_s}{dz} = \frac{2id_{eff}\omega_s}{n_s c} E_{SFG} E_p^* \exp(-i\Delta kz) \\ \frac{dE_{SFG}}{dz} = -\frac{2id_{eff}\omega_{SFG}}{n_{SFG} c} E_p E_s \exp(i\Delta kz) \end{cases}, \quad (S4)$$

where $d_{eff}$ is the effective nonlinear coefficient, $\Delta k = k_{SFG} - k_s - k_p + 2\pi/\Lambda$ is the phase mismatch and $\Lambda$ is the poling period. We assume that the crystal is nonabsorptive for all three waves and that the phase mismatch $\Delta k = 0$. By integrating over the crystal length, the SFG beam amplitude is

$$E_{SFG}(r,L/2) = -\frac{2id_{eff}\omega_{SFG}}{n_{SFG}c}\int_{-L/2}^{L/2} E_p(r,z)E_s(r,z)dz, \quad (S5)$$

where $L$ is the crystal length. The SFG power can then be expressed as

$$P_{SFG} = 2\varepsilon_0 c n_{SFG}\int_0^{2\pi} d\theta \int_0^{\infty} |E(r,L/2)|^2 r\, dr. \quad (S6)$$

By inserting Equations (S2) and (S3) into Equation (S6), and integrating over the cross-section of the beam, we derive the following expression for the SFG power:

$$P_{SFG} = \frac{16\pi^2 d_{eff}^2 P P_s L 2^l}{\varepsilon_0 c n_s n_{SFG}\lambda_{SFG}^2 \lambda_p} h(l,\xi). \quad (S7)$$

Here, $\xi = L/2Z_{0p}$ is the focusing parameter, and $h(l,\xi)$ is defined as

$$h(l,\xi) = \frac{1}{\xi}\int_{-\xi}^{\xi} dx \int_{-\xi}^{\xi} dy \frac{(1+ix)^l (1-iy)^l}{\{(1+ix)(1-iy)[2+\frac{i(x-y)}{\beta}]+\alpha(1+\frac{ix}{\beta})(1-\frac{iy}{\beta})[2+i(x-y)]\}^{l+1}} , \quad (S8)$$

where $\alpha = \omega_{0s}^2 / \omega_{0p}^2$ and $\beta = Z_{0s}/Z_{0p} = n_s \lambda_p \omega_{0s}^2 / n_p \lambda_s \omega_{0p}^2 = n_s \lambda_p \alpha / n_p \lambda_s$. When $l=0$, Equation (S7) is reduced to

$$P_{SFG} = \frac{16\pi^2 d_{eff}^2 P P_s L}{\varepsilon_0 c n_s n_{SFG} \lambda_{SFG}^2 \lambda_p} h(0,\xi) . \quad (S9)$$

Then the maximum pump power required to reach unity conversion quantum efficiency is

$$P_{max} = \frac{\varepsilon_0 c n_s n_{SFG} \lambda_p \lambda_s \lambda_{SFG}}{16\pi^2 d_{eff}^2 L h(0,\xi)} , \quad (S10)$$

We can use Equation (S10) to guide the design of the SFG cavity and explain the experimentally measured quantum efficiency in the up-conversion process.

The measured power circulating in our cavity is 22.3 W, and the experimentally estimated maximum pump power from Equation (8) is 177 W. The theoretically estimated power using Equation (S10) is 170 W, based on the following parameters: $\alpha = 1$, $\xi = 0.3$ and $h(0,\xi) = 0.242$.

**Explanation of the derivation of equation (10) and the method used for measuring the interference**

The function of the HWP before the SFG module is used to transform the polarization of the state in equation (9). The H and V polarization in equation (9) are transformed as following:

$$\begin{cases} |H\rangle \rightarrow \frac{1}{\sqrt{2}}(|H\rangle + |V\rangle) \\ |V\rangle \rightarrow \frac{1}{\sqrt{2}}(|H\rangle - |V\rangle) \end{cases} , \quad (S11)$$

Therefore equation (9) is transformed to

$$|\Phi\rangle_{in} \rightarrow |\Phi\rangle = \frac{1}{2}(|l\rangle + e^{i\theta}|-l\rangle) \otimes |H\rangle + \frac{1}{2}(|l\rangle - e^{i\theta}|-l\rangle) \otimes |V\rangle , \quad (S12)$$

Only the vertical polarization part of equation (S12) can matching the phase matching condition in the up-conversion process, therefore the up-conversion projected the states in equation (S12) into the vertical polarization. The final states of the up-conversion OAM state is

$$|\Phi\rangle_{out} = \frac{1}{\sqrt{2}}(|l\rangle - e^{i\theta}|-l\rangle) \otimes |V\rangle . \quad (S13)$$

Equation (S13) differs with equation (10) in a minus sign, this minus sign is absorbed into the phase $\theta$. The superposition states in equation (S13) has intensity distribution with 2l maximum in the azimuthal direction, the angle separation period is $\pi/l$, which is petal-like. We use $l=2$ as an instance to show how to measure the interference in the main text(see figure S1 below). We first filtering out the red dot labeled petal using a pin hole and coupling the transmitted beam to single mode fiber, then we rotate the HWP at the input port of the Sagnac interferometer, as the phase $\theta$ is dependent on the axes position of the HWP. By rotating the HWP, the interference pattern will rotate gradually, the overlap of the petal with the pin hole changed periodically. When the phase $\theta$ changes $2\pi$, the interference pattern rotates $\pi/2$, which is the angle period in the azimuthal direction.

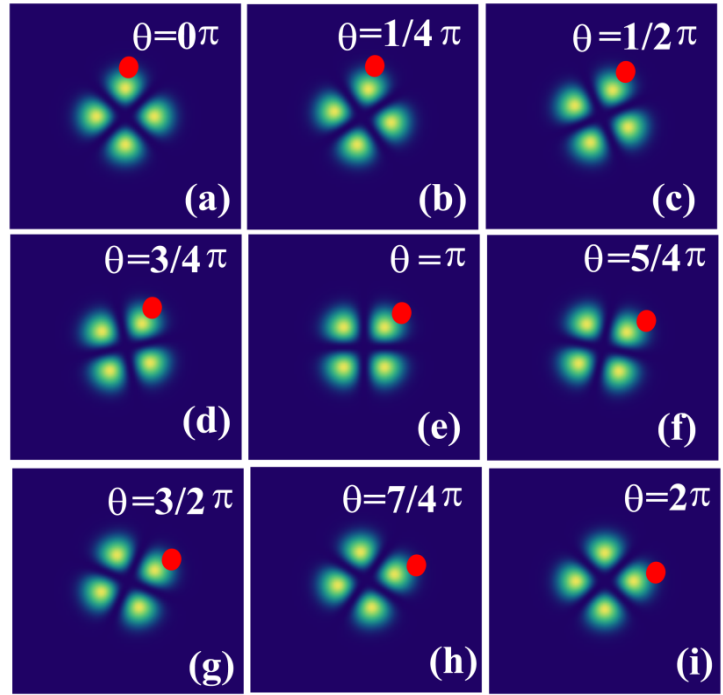

Figure S1. The rotation of the red dot labeled petal as a function of the relative phase $\theta$ in the superposition states equation (S13).

**Estimation of single photon conversion efficiency and further optimizations**

The single photon conversion efficiency for a Gaussian pump can be estimated from Figure 3(e), and the herald rate of our up-conversion system is $1.00\times10^{-3}$. The losses of the signal photon after emission from the PPKTP crystal include losses from the fiber coupling efficiency of FC1 (0.25), the transmission loss before the input face of the SFG crystal (0.80), the filtering loss of the up-converted photons (0.80), the fiber coupling efficiency of FC3 (0.50), the bandwidth mismatch loss (the SFG bandwidth is 1 nm, while the signal photon bandwidth is 2.44 nm, so only 0.41 of each signal photon is up-converted), and losses from the single photon detection efficiency of APD2 (0.50). When we take all these losses into account, the internal single photon conversion quantum efficiency of the Gaussian mode is 0.061, and the conversion efficiency for mode $l=1$ is 0.02.

Future possible improvements to the present setup are as follows: the conversion efficiency can be increased by using a longer crystal and by choosing the optimal transmittance for the input coupling mirror. The use of a longer crystal for the SPDC photon source will enable the bandwidth match between the SPDC photon and the SFG; the noise photons of the up-converter coming from the SPDC and the SRS can be reduced dramatically by using a pump with a longer wavelength and by using narrower filters.

**Noise analysis**

The two primary noise sources in quantum frequency conversion are SPDC and spontaneous Raman scattering (SRS). The noise photons generated by these two processes within the acceptance bandwidth of the SFG crystal will be up-converted, and the fidelity of the up-conversion process will be reduced. SPDC only generates photons with frequencies that are lower than the pump frequency; this photon noise is a result of random duty-cycle errors that are inherent in the fabrication of quasi-phase-matching gratings by periodic poling. The SPDC noise is a statistically white noise floor for all wavelengths longer than the pump wavelength within the transparent window of the crystal. However, SRS can generate either red- or blue-shifted photons from the pump as Stokes or anti-Stokes sidebands, which are generated with an intensity ratio that is determined by a Boltzmann factor related to the thermal occupation of photons. By using a longer pump wavelength, the noise arising from the SPDC of the pump can be avoided and the SRS noise can be reduced dramatically. In addition, a narrower filter bandwidth will also result in a lower noise ratio. In the experiments presented here, the filter bandwidth is 40 nm and the pump wavelength is shorter than the signal wavelength, which limits the experimental performance. Better experimental results are expected through the use of a filter with a narrower bandwidth and a pump operating at a longer wavelength.